\documentclass[12pt]{JHEP3}

\usepackage{amsmath}
\usepackage{amsfonts}
\usepackage{amssymb}
\usepackage{graphicx}
\usepackage{epstopdf}
\usepackage{bbm}

\def\msbar{\overline{\mathrm{MS}}}

\title{Higher representations on the lattice: perturbative studies
}

\author{Luigi Del Debbio \\ 
        SUPA, School of Physics and Astronomy, University of Edinburgh \\
	Edinburgh EH9 3JZ, Scotland~\footnote{Permanent address.}\ , {\rm and} \\
        Isaac Newton Institute for Mathematical Sciences\\
        20 Clarkson Road, Cambridge CB3 0EH, UK\\
	E-mail: \email{luigi.del.debbio@ed.ac.uk} 
} 
\author{Mads T. Frandsen \\
        University of Southern Denmark, \\
        Campusvej 55, DK-5230 Odense M, Denmark, {\it and}\\
        Niels Bohr Institute, 
        Blegdamsvej 17, DK-2100 Copenhagen, Denmark\\
	E-mail: \email{toudal@nbi.dk} 
}
\author{Haralambos Panagopoulos \\
	Department of Physics, University of Cyprus, Kallipoleos 75, P.O.B. 20537\\
	Nicosia CY-1678, Cyprus \\ 
	E-mail: \email{haris@ucy.ac.cy} 
} 
\author{Francesco Sannino \\
        University of Southern Denmark, \\
        Campusvej 55, DK-5230 Odense M, Denmark, \\
	E-mail: \email{sannino@fysik.sdu.dk} 
}

\abstract{We present analytical results to guide numerical simulations
  with Wilson fermions in higher representations of the colour
  group. The ratio of $\Lambda$ parameters, the additive
  renormalization of the fermion mass, and the renormalization of
  fermion bilinears are computed in perturbation theory, including
  cactus resummation. We recall the chiral Lagrangian for the
  different patterns of symmetry breaking that can take place with
  fermions in higher representations, and discuss the possibility of
  an Aoki phase as the fermion mass is reduced at finite lattice
  spacing.}


\preprint{NI08005}

\begin{document}

\section{Introduction}
\label{sec:intro}

The nonperturbative dynamics of asymptotically free gauge theories
with matter fields transforming according to higher dimensional
representations of the underlying gauge group is a topic of current
research interest.  Physics beyond the Standard Model (BSM) offers a
new arena for the use of these theories, which are expected to develop
a nonperturbative infrared fixed point for a very low number of
flavors \cite{Sannino:2004qp,Dietrich:2006cm}. Because they minimize
the tension with the electroweak precision constraints, some of these
theories are excellent candidates for the dynamical breaking of the
electroweak symmetry of walking technicolor type, which were first
introduced in
Refs.~\cite{Holdom:1984sk,Holdom:1983kw,Eichten:1979ah,Holdom:1981rm,Yamawaki:1985zg,Lane:1989ej}. Since
the number of flavors needed to get near the conformal fixed point is
small, the associated models have been termed {\it minimal walking
  technicolor} \cite{Sannino:2004qp,Dietrich:2006cm,Foadi:2007ue}.  By
walking one refers to the fact that the running coupling decreases
much more slowly with the reference energy scale than in the case of
QCD--like theories. Yet, another interesting physical application of
the study of the phase diagram of strongly coupled theories is to
provide the theoretical landscape underlying the {\it unparticle}
physics world~\cite{Georgi:2007ek,Georgi:2007si}. The theory landscape
was provided in Ref.~\cite{Ryttov:2007sr} where it was shown that the
fraction of asymptotically free gauge theories developing an infrared
fixed point is quite large. Studying their phase diagram is a
fundamental step if these theories aspire to become realistic
candidates for BSM physics. Insight into the phase diagram of such
theories has been recently provided by a proposed all-order beta
function for any number of colours and for any
representation~\cite{Ryttov:2007cx}.  Moreover, in the limit of a
large number of colours, the planar orientifold planar equivalence
relates theories with fermions in higher representations to
supersymmetric theories~\cite{Armoni:2003gp}. It provides interesting
predictions that deserve nonperturbative
investigations~\cite{Armoni:2003fb,Armoni:2003yv}. The necessity to
study the large-$N$ limit makes these theories more expensive to study
numerically. Finally, understanding the strong dynamics that governs
the low--energy behaviour of such theories is an interesting problem
{\it per se}.

Lattice is a privileged tool for exploring the nonperturbative
dynamics of strongly interacting theories, but Monte Carlo
simulations of these theories can capture the interesting dynamical
features only if the full fermion determinant is taken into account in
the Boltzmann weight used for generating gauge configurations. So far
only limited experience has been gathered from numerical simulations
with dynamical fermions beyond
QCD~\cite{Campos:1999du,Catterall:2007yx}. In the light of recent
algorithmic progresses in simulating quantum field theories with
dynamical
fermions~\cite{Hasenbusch:2001ne,Hasenbusch:2002ai,Luscher:2005rx,DelDebbio:2005qa,Urbach:2005ji,Clark:2006fx},
numerical studies with fermions in higher--dimensional representations
are now a realistic target (see Ref.~\cite{Catterall:2007yx} for early
work in this direction). As a preliminary work, which provides
guidance for large--scale simulations, we present here an
investigation of the space of bare couplings by analytical tools, such
as perturbation theory and chiral Lagrangians.

Perturbative results are useful to understand the behaviour of the
lattice theory as the continuum limit is approached: on one hand they
provide a connection between the lattice results and their continuum
counterparts; on the other hand they offer some quantitative support
in choosing the bare parameters in the early stages of numerical
simulations. Precision studies in QCD have shown sizable discrepancies
between perturbative and nonperturbative computations at the values of
the bare parameters that are currently accessible. Assessing the
accuracy of perturbation theory for theories with fermions in
higher--dimensional representations is beyond the scope of this paper
and will be deferred to future publications. Instead we shall
supplement perturbative calculations with sensible assumptions, that
we discuss below, in order to dictate the choice of the values of the
bare parameters for first numerical investigations.

In this work perturbation theory is used to determine the dependence
of the lattice spacing on the bare lattice coupling, the ratio of
lambda parameters $\Lambda_{\msbar}/\Lambda_\mathrm{lat}$, the
critical value of the bare mass $m_c$ and critical hopping parameter
$\kappa_c$, and the renormalization 
constants for fermionic bilinears. The gauge action considered is the
simple plaquette action, and the fermion action is the unimproved
Wilson action. This simple choice provides a concrete example for
performing the perturbative calculations, and matches the existing,
and forthcoming numerical simulations. Four specific examples of
lattice theories with Wilson fermions are compared below. Results for
quenched QCD (QCD0), and for QCD with two flavours of dynamical
fermions (QCD2) are known both in perturbation theory, and from
non--perturbative computations. They are briefly summarized in this
work in order to set the framework for our computations, and to assess
the accuracy of perturbation theory. We then present the
generalization of the perturbative calculations to arbitrary
representations, and analyze in detail their implications for two
theories that are good candidates for BSM phenomenology, namely the
SU(3) gauge theory with $n_f=2$ flavours in the two--index symmetric
representation (T1), and the SU(2) gauge theory with two flavours in
the two--index symmetric representation (T2). 

The chiral Lagrangian describing the dynamics of the light Goldstone
bosons is analyzed in order to clarify the structure of the phase
diagram that is likely to be revealed by numerical simulations for
small quark masses. We discuss in particular the theories with
fermions in higher representations introduced above. Note that the
same approach can be readily applied to other theories, like
e.g. gauge theories with fermions in the two--index antisymmetric
representations, that are interesting for numerical tests of the
planar orientifold equivalence. Even though we do not discuss these
theories explicitly here, our conclusions can be specialized in a
straightforward manner.

The paper is organized as follows. In Sect.~\ref{sec:a} we discuss the
perturbative results describing asymptotic scaling, we compute the
ratio of $\Lambda$ parameters, and discuss the approach of the
continuum limit. Sect.~\ref{sec:b} reports some useful results at
one--loop in perturbation theory. We first consider the
renormalization of the bare mass for Wilson fermions; the critical
value of the hopping parameter $\kappa_c$ is computed both up to two
loops, and using the so--called cactus dressing to resum a particular
class of tadpole diagrams~\cite{Panagopoulos:1998xf}. Similarly we
present results for the renormalization constants for fermion
bilinears. Finally in Sect.~\ref{sec:d} we discuss the form of the
chiral Lagrangian that describes the low--energy dynamics in theories
with fermions in arbitrary representation and the possible phase
structure as the quark mass is lowered at finite lattice spacing.

Numerical simulations of the theories with fermions in higher
representations are deferred to further publications. 

\section{Scaling}
\label{sec:a}

\subsection{Ratio of $\Lambda$ parameters}
The $\beta$ function encodes the dependence of the lattice spacing $a$
on the bare coupling constant $g_0$. In mass--independent
renormalization schemes, the lattice spacing is uniquely determined by
the bare coupling, according to the renormalization group equation:
\begin{equation}
\label{eq:betalat}
\beta_\mathrm{lat}(g_0)=-a \left. \frac{\partial g_0}{\partial a} \right |_{g_R~\mathrm{fixed}},
\end{equation}
where the subscript $\mathrm{lat}$ refers to the lattice scheme which is being considered here.

For a generic gauge theory, with gauge group SU($N$) and $n_f$
fermions in a given representation $R$ of the colour group, the
two--loop computation in perturbation theory yields the familiar
expression:
\begin{eqnarray}
\beta_\mathrm{lat}(g_0)&=&-\beta_0 g_0^3 - \beta_1 g_0^5 + O(g_0^7) \\
\beta_0&=&\frac{1}{(4 \pi)^2} \left[ \frac{11}{3} C_2(A) - \frac43 T_R n_f \right]\label{eq:beta0} \\
\beta_1&=&\frac{1}{(4 \pi)^4} \left[ \frac{34}{3} C_2(A)^2 - \frac{20}{3} C_2(A) T_R n_f - 4 C_2(R) T_R n_f \right],
\end{eqnarray}
where $T_R$ yields the normalization of the generators, and $C_2(R)$
is the quadratic Casimir, both in the representation $R$. Factors of
$C_2(A)$, the quadratic Casimir in the adjoint representation, arise
because of gluon loops, and do not change as the fermionic
representation is varied. Note that the first two coefficients of the
$\beta$ function are universal and depend neither on the
regularization nor on the
renormalization scheme. For $N=3$ and fermions in the fundamental
representation, the expressions above reduce to the usual values of
$\beta_0,\beta_1$. Tables for the group--theoretical factors for the
representations considered in this work are reported in
App.~\ref{app:a}.

The asymptotic behaviour of $a(g_0)$ is obtained by integrating
Eq.~(\ref{eq:betalat}), and the scale $\Lambda_\mathrm{lat}$ is the
integration constant that appears in this procedure. Following the
notation in Ref.~\cite{Weisz:1980pu}, we write:
\begin{eqnarray}
a(g_0) \Lambda_\mathrm{lat}&=&
\exp\left[-\int^{g_0}\frac{dg^\prime}{\beta_\mathrm{lat}(g^\prime)}\right]\nonumber\\
&=&\exp\left[-1/(2\beta_0 g_0^2)\right] 
\left(\beta_0 g_0^2\right)^{-\beta_1/(2 \beta_0^2)} \left[1+O(g_0^2)\right].
\end{eqnarray}

The ratio of $\Lambda$--parameters defined in different
renormalization schemes is obtained from the one--loop relation
between the coupling
constants~\cite{Dashen:1980vm,Hasenfratz:1980kn,Kawai:1980ja,Weisz:1980pu}. In
particular the running coupling in the $\msbar$ scheme is related to
the lattice bare coupling via:
\begin{equation}
\label{eq:alpharel}
g_{\msbar}(\mu)=\left\{1+\sum_{l=1}^{\infty} {\mathcal Z}^{(l)}(\mu a,\lambda_0) g_0^{2l}\right\}^{-1/2} g_0,
\end{equation}
where $\lambda_0$ is the bare gauge--fixing parameter in the lattice
formulation.  This relation can be obtained {\it e.g.} using the background
field technique~\cite{Abbott:1980hw,Ellis:1983af,Luscher:1995vs}. The
first coefficient ${\mathcal Z}^{(1)}(\mu a,\lambda_0)$ for matter
fields in the fundamental representation can be found in the
literature, see {\it e.g.}\ Ref.~\cite{Kawai:1980ja,Weisz:1980pu}. It
has the generic form:
\begin{equation}
\label{eq:Zg1}
\left.{\mathcal Z}^{(1)}(\mu a,\lambda_0)\right|_{\lambda_0=1}=\beta_0 \log(a^2 \mu^2)+l_0,
\end{equation}
and the ratio of $\Lambda$--parameters is obtained from the coefficients
in Eq.~(\ref{eq:Zg1}) as:
\begin{equation}
\Lambda_\mathrm{lat}/\Lambda_{\msbar}=\exp\left[l_0/(2\beta_0)\right].
\end{equation}
Having already written the coefficient $\beta_0$ for a generic
representation in Eq.~(\ref{eq:beta0}), the expression for the finite
part of the one--loop contribution in Eq.~(\ref{eq:alpharel}), $l_0$,
is the only ingredient needed in order to convert the
$\Lambda$--parameter. The coefficient $l_0$ is obtained by inspecting
the one--loop diagrams that contribute to Eq.~(\ref{eq:Zg1}). The
group--theoretical factors need to be changed in order to take into
account the new fermionic representation, while the numerical factors
that arise from the integration over the lattice momenta remain
unchanged. For a generic representation $R$, we obtain:
\begin{eqnarray}
\label{eq:l0}
l_0&=&\frac{1}{(8 \pi^2)}\left[-2\pi^2 \; C_2(F) - 3.54958342046\;
  C_2(A) + \nonumber \right. \\
   && \left. + 1.057389936\; T_R n_f \right],
\end{eqnarray}
where the only dependence on the fermionic representation is encoded
in the last term in the sum on the RHS of Eq.~(\ref{eq:l0}).  Explicit
values for some representations of interest are summarized in
Tab.~\ref{tab:coeffs}. The well--known values for the quenched SU(2)
and SU(3) theories, and for QCD with two flavours of fundamental
fermions are reported in order to show explicitly the differences in
the perturbative coefficients as we introduce matter in higher
representations. As a non--trivial check, we can specialize
Eq.~(\ref{eq:l0}) to the case of ${\mathcal N}=1$ SYM, which
corresponds to $n_f=1/2$ flavour of fermions in the adjoint
representation. Using the group--theoretical factors reported in
Tab.~\ref{table1} in Appendix~\ref{app:a}, our expression reproduces
Eq.~(24) in Ref.~\cite{Montvay:1997uq}. 
\TABLE[ht]{
  \begin{tabular}[h]{llll}
    Representation & $\beta_0$ & $\beta_1$  & $l_0$ \\
    \hline
    SU(2), $n_f=0$        & 0.0464389 &  0.00181793 & -0.277412   \\ 
    SU(3), $n_f=0$        & 0.0696583 &  0.00409035 & -0.468201   \\ 
    SU(2), $n_f=2$, fund  & 0.0612149 &  0.00307445 & -0.454809   \\ 
    SU(3), $n_f=2$, fund  & 0.0612149 &  0.00307445 & -0.454809   \\ 
    SU(3), $n_f=2$, 2S    & 0.0274412 & -0.00259323 & -0.401241   \\ 
    SU(2), $n_f=2$, 2S    & 0.0126651 & -0.00160406 & -0.223844   \\ 
    \hline
  \end{tabular}
  \caption{Perturbative coefficients appearing in one-loop perturbative computations.}
  \label{tab:coeffs}
}

The ratios of $\Lambda$--parameters are easily obtained from the
values in Tab.~\ref{tab:coeffs}.  The coefficients in the first lines
of the table reproduce the known results:
\begin{eqnarray}
  \label{eq:oldreslambda}
  \left.\Lambda_{\msbar}/\Lambda_\mathrm{lat}\right|_{\mathrm{SU(2)},
    n_f=0} &=& 19.82, \\
  \left.\Lambda_{\msbar}/\Lambda_\mathrm{lat}\right|_{\mathrm{SU(3)},
    n_f=0} &=& 28.81, \\
  \left.\Lambda_{\msbar}/\Lambda_\mathrm{lat}\right|_{\mathrm{SU(3)},
    n_f=2, \mathrm{fund}} &=& 41.05,   
\end{eqnarray}
see {e.g.}\
Refs.~\cite{Hasenfratz:1980kn,Kawai:1980ja,Weisz:1980pu}. The last two
lines yield the new results for the higher representations that we
want to consider in this work:
\begin{eqnarray}
  \label{eq:lambdarat}
  \left.\Lambda_{\msbar}/\Lambda_\mathrm{lat}\right|_{\mathrm{SU(3)}, n_f=2, 2S}
  &=& 1469.59 \\
  \left.\Lambda_{\msbar}/\Lambda_\mathrm{lat}\right|_{\mathrm{SU(2)}, n_f=2, 2S}
  &=& 6884.36. 
\end{eqnarray}
Values for other representations can be easily deduced from the
formulae above. Including two flavours of fermions in higher
representations induces large variations in the ratios of $\Lambda$
parameters. This is at odds with the results for fermions in the
fundamental representation, where adding the effect of fermion loops
yields a much smaller variation of the ratio. However the large values
obtained for fermions in two--index representations can be understood
by rewriting the ratio $\beta_0/2 l_0$ in a way which makes the $1/N$
scaling explicit:
\begin{equation}
  \label{eq:Nscal}
  \frac{l_0}{2\beta_0}\simeq -3.65978 \cdot
  \frac{1-0.0787969 T_R \frac{n_f}{N} -0.735484\frac{1}{N^2}}
  {1-\frac{4}{11} T_R \frac{n_f}{N}}.
\end{equation}
In Eq.~(\ref{eq:Nscal}) we can easily recognize the contributions
$O(n_f/N)$ from fermion loops, and the contributions from the
non--planar gluonic diagrams of $O(1/N^2)$. For fer\-mi\-ons in the
fundamental representation $T_R=1/2$, and therefore the fermion
determinant yields corrections that are suppressed by $n_f/N$. For
fermions in 2--index representations, the normalization of the
generators is such that $T_R\sim O(N)$, and hence the contribution
from the fermion determinant is of the same order of the gluon
contribution, both in the numerator and the denominator, and therefore
large variations are found with respect to the pure gauge theory.

\subsection{Perturbative and nonperturbative scaling}
The perturbative results obtained in the previous subsection can be
used to sketch the scaling of the lattice spacing for theories in
higher representations, being well aware of the limitations of
perturbation theory. The nonperturbative scaling of the lattice
spacing has been carefully studied for QCD, both in the quenched
approximation, and for the theory with dynamical quarks in the
fundamental representation.  For the latter theories, the accuracy of
perturbative estimates can be assessed by comparing numerical and
analytical results. As we shall see below, perturbation theory in QCD
does not yield an accurate description of the scaling of physical
quantities. Therefore, any result obtained in this framework is bound to
be approximate and should be used mostly as a guide for forthcoming
numerical simulations.

In order to relate more easily to the notation used in numerical
simulations, let us introduce the lattice coupling
$\beta=2N/g_0^2$. We will henceforth use $\beta$ to indicate the bare
lattice coupling, unless explicitly stated. The asymptotic scaling
formula reported in the previous subsection yields the value of the
lattice spacing in physical units as a function of $\beta$:
\begin{equation}
\label{eq:latpert}
\frac{a^{-1}(\beta)}{\Lambda_{\msbar}}=
\left(\frac{\Lambda_\mathrm{lat}}{\Lambda_{\msbar}}\right) 
\exp\left[\frac{\beta}{4 N \beta_0}\right]
\left( 2 N \beta_0/\beta\right)^{\beta_1/(2 \beta_0^2)}
\left[1 + O(1/\beta)\right]
.
\end{equation}
Having computed the ratio of $\Lambda$ parameters, the only
input that is required is the value of $\Lambda_{\msbar}$. 

For the SU(3) pure gauge theory, Fig.~\ref{fig:qcd0.scal} displays the
prediction for the lattice spacing $a$ in physical units [fm],
computed from two--loop perturbation theory using the input from
Ref.~\cite{Capitani:1997mw,Capitani:1998mq}; the curve is compared to
the interpolation of the nonperturbative data presented
in~\cite{Guagnelli:1998ud}. The error band in the figure is simply the
error that is obtained from propagating the error in the determination
of $\Lambda_{\msbar}$ to the value of $a(\beta)$.  As shown by the
plot, the perturbative prediction in bare perturbation theory
underestimates the actual lattice spacing by 30--50\% at the values of
$\beta$ between 5.8 and 6.2, where most simulations have been
performed so far. The two computations agree at large values of
$\beta$, as expected when the continuum limit is approached.
\FIGURE[h]{
\includegraphics[width=0.8\textwidth]{FIGS/QCD0.eps}
\caption{Comparison of the lattice scaling in physical units predicted
  from perturbation theory with the nonperturbative results obtained
  from numerical simulations. The theory is pure gauge SU(3). The blue
  (respectively red) curve represents the perturbative
  (resp. nonperturbative) estimate of the lattice spacing in fm as a
  function of the lattice bare coupling $\beta$. The error on the
  perturbative estimate comes from the error in the determination of
  $\Lambda_{\msbar}$. The red curve is an interpolation of the
  nonpertubative determination of the lattice spacing. }
\label{fig:qcd0.scal}
}

For the SU(3) theory with two flavours of Wilson fermions, the
perturbative prediction is obtained using the $\Lambda$ parameter
computed in Ref.~\cite{Sommer:2006wx,DellaMorte:2004bc}; it can be
compared to the value of the lattice spacing recently computed for two
values of $\beta$ in Ref.~\cite{DelDebbio:2006cn}. Again, in the range
that is accessible to current simulations, the perturbative estimate
is smaller by approximately 30\%--40\%. Such large deviations between
lattice bare perturbation theory and nonperturbative results are very
well-known to lattice QCD practitioners~\cite{Lepage:1992xa}. They are
reported here in order to have a concise summary of the results in
QCD, before moving into new territories. 

Figure~\ref{fig:qcd0.alambda} shows the perturbative estimate for the
value of the dimensionless quantity $a(\beta) \Lambda_{\msbar}$ for
QCD with $n_f=0,2$ in the same range of $\beta$; for the values of
$\beta$ used in current simulations, one can see that $1/(a
\Lambda_{\msbar})\approx 20$. Given that the hadron masses in QCD turn
out to be of the order of $\Lambda_{\msbar}$, the value of $1/(a
\Lambda_{\msbar})$ is such that lattice artefacts are small, while
sufficiently large physical volumes can be reached on lattices that
have 20--30 points in each direction. Lattice simulations for theories
beyond QCD need to identify a similar regime in order to avoid large
lattice artefacts and/or large finite volume effects.
\FIGURE[h]{
\includegraphics{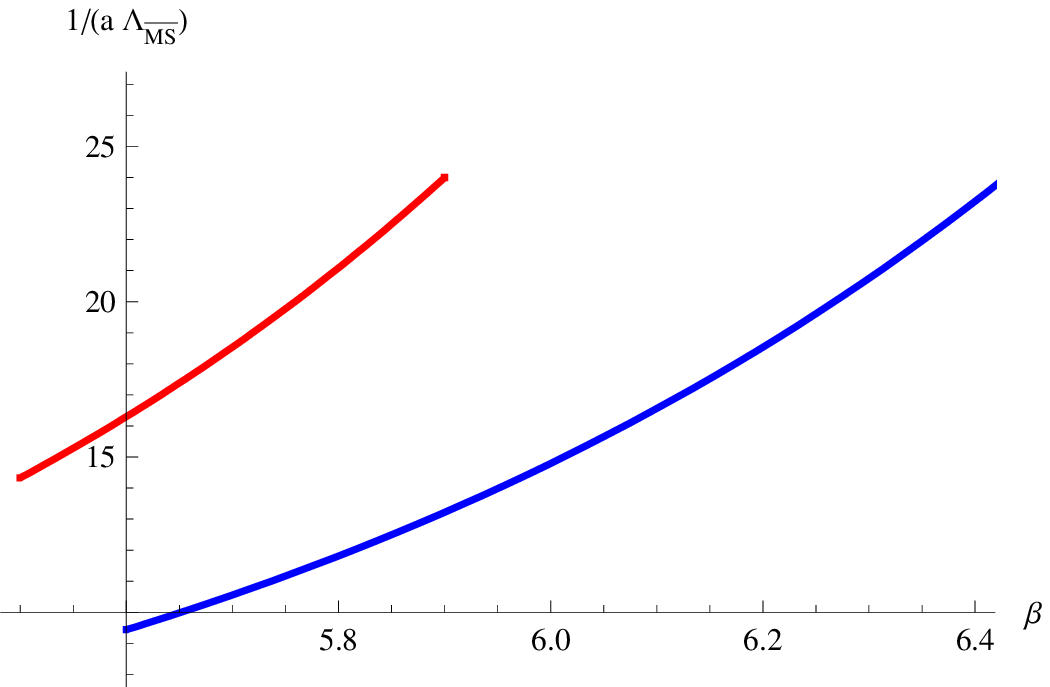}
\caption{The perturbative result for the dimensionless quantity
  $a(\beta) \Lambda_{\msbar}$ as a function of $\beta$ for pure gauge
  SU(3) (blue line), and SU(3) with $n_f=2$ flavours of Wilson
  fermions (red line). }
\label{fig:qcd0.alambda}
}

Perturbative results can be used for the theories with fermions in
higher representations in order to arrive at an educated guess for the
value of $\Lambda_{\msbar} a(\beta)$ from perturbative scaling,
provided a few hypotheses are made in order to identify the relevant
energy scales.  As already mentioned in the introduction, a
near--conformal behaviour is expected in the theories T1 and T2. The
dependence of the coupling on the lattice spacing in these theories is
characterized by two different regimes. At high energies the theories
are asymptotically free, and therefore we expect the usual logarithmic
running of the renormalized coupling. However, as the energy scale is
decreased, it should reach a value, which we denote
$\Lambda_{\mathrm{w}}$, where the coupling starts to ``walk'',
i.e. where the coupling is only weakly dependent on the cutoff.  The
walking behaviour should extend for several orders of magnitude, until
a lower scale $\Lambda_{\mathrm{IR}}$ where the coupling starts
running again. Phenomenologically relevant models would favour a ratio
$\Lambda_{\mathrm{w}}/\Lambda_{\mathrm{IR}}\geq
10^3$~\cite{Dietrich:2006cm}. However, it should be noted that the
running of the coupling constant is scheme dependent, and therefore
this ratio should only be taken as an indicative value. 

The perturbative values of $\Lambda_{\msbar} a(\beta)$ as a function
of $\beta$ are reported in Fig.~\ref{fig:t1scal} for the theories T1
and T2. If we assume that $\Lambda_{\msbar}\sim\Lambda_\mathrm{IR}$,
and if we further require that $\Lambda_{\mathrm{w}} a(\beta) \approx
0.1$, then numerical simulations of phenomenologically relevant models
would require $\Lambda_{\msbar} a(\beta) \leq 10^{-5}$. The range of
$\beta$ in the figure is chosen to yield values of $\Lambda_{\msbar}
a(\beta)$ that saturate the inequality above; if the ratio between the
typical hadron masses turn out to be large in units of
$\Lambda_{\msbar}$, so that $m_\mathrm{had} a \simeq 1$, then higher
values of $\beta$ would be needed in order to keep both lattice
artefacts and finite volume effects under control. More quantitative
information can only be obtained from numerical simulations.
\FIGURE[h]{
  \includegraphics[width=0.45\textwidth]{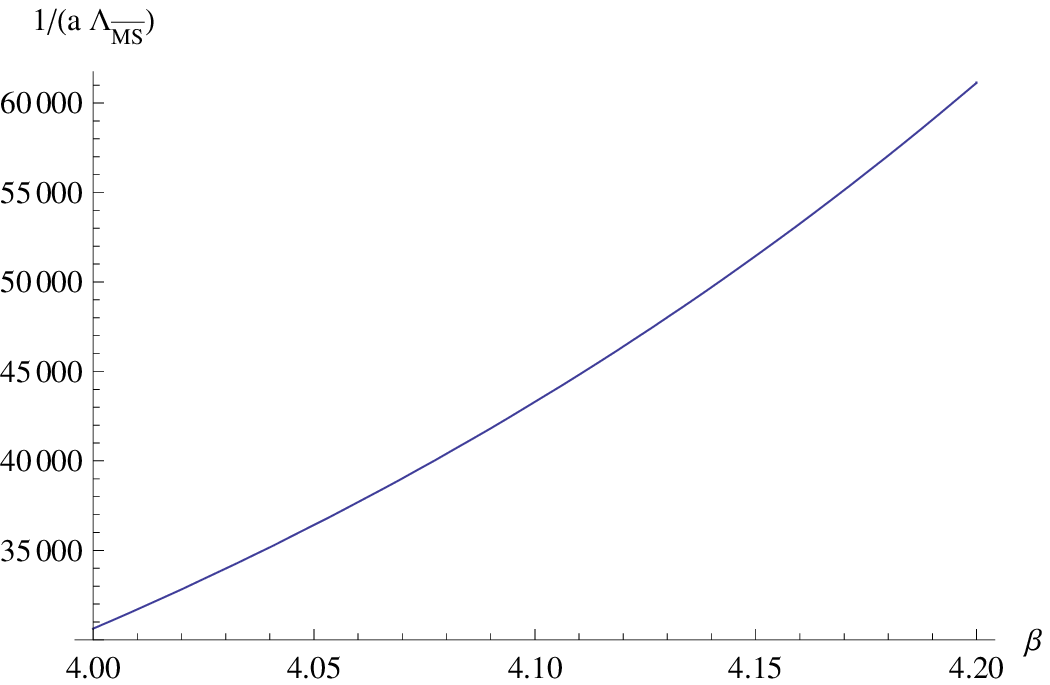}
  \includegraphics[width=0.45\textwidth]{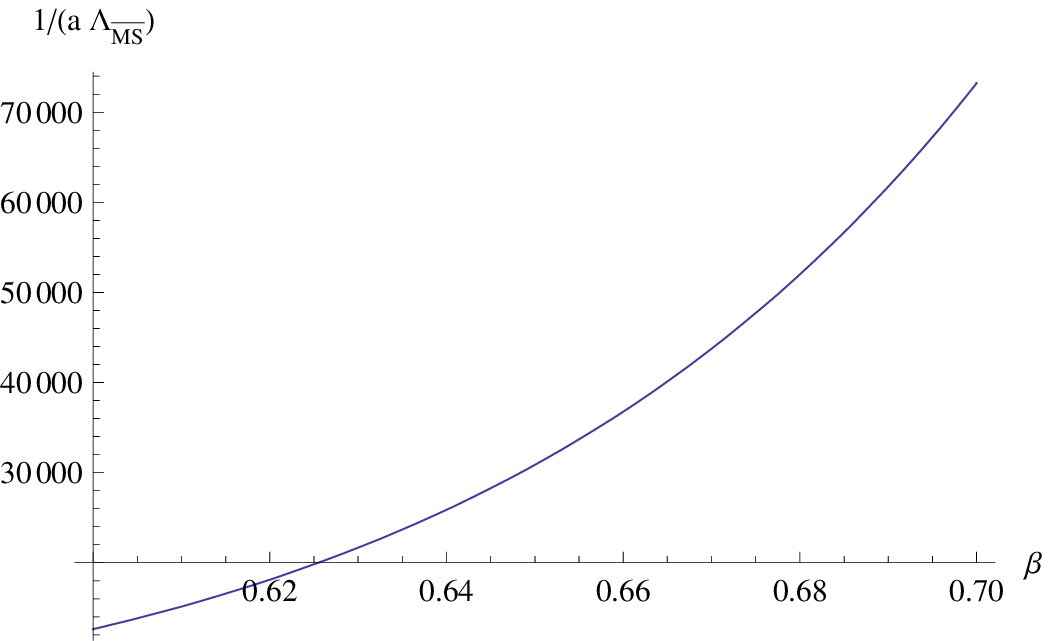}
  \caption{The lattice spacing in unit of $\Lambda_{\msbar}$ as a
    function of $\beta$ for the theory T1 (left), and T2 (right).}
  \label{fig:t1scal}
}
As already seen above, in numerical simulations of QCD nonperturbative
scaling does deviate from the two--loop predictions by up to 50\%. We
should therefore take these perturbative results with a grain of
salt. 

It is worthwhile to emphasize that, while we have expressed everything
in units of $\Lambda_{\msbar}$, the absolute value of the scale for
these new theories is not known a priori, and can only be determined
by computing some physical dimensionful quantity. An obvious candidate
would be the decay constant of the technipion in the chiral limit,
which is related to the Higgs {\it vev} and can be estimated to be
$F\approx 250~\mathrm{GeV}$.

Nonperturbative results that could highlight the near--conformal
behaviour can be obtained by Monte Carlo renormalization group
methods~\cite{Bowler:1984hv,Heller:1985xj}. However these methods
require to vary the lattice cutoff over a large interval, while
simultaneously keeping the finite--size errors under control. If the
physical scales of interest are well separated, lattice simulations
require a very fine resolution, i.e. a large number of lattice points,
that may not be accessible with present--day computing resources.

The finite--volume schemes introduced in Ref.~\cite{Luscher:1991wu}
provide an elegant solution to this problem; the renormalization scale
is identified with the inverse of the linear lattice size, and the
evolution of the renormalized coupling is computed in steps,
changing the scale $\mu$ by factors of 2 in each step. The variation
of the coupling is summarized in the step--scaling function, which
yields a precise determination of the nonperturbative beta
function. Recent results for the theory with $n_f=12$ fermions in the
fundamental representation show that this is a promising way to
study the problem~\cite{Appelquist:2007hu}. 

\section{Perturbative renormalization}
\label{sec:b}

In order to perform numerical simulations, preliminary estimates of
the critical mass and of the renormalization of fermion bilinears are
needed. This section summarizes some useful computations at one and
two loops in perturbation theory for fermions in generic
representations, that can be used to guide preliminary lattice
studies. We will consider here the theory defined on the lattice, with
Wilson action for the fermions, and simple plaquette action for the
gauge fields. In the gauge action, the link variables are always
SU($N$) matrices in the fundamental representation, while in the
fermionic part of the action, the covariant derivatives are defined
through the link variables in the generic representation. Feynman
rules for perturbative calculations are easily generalized.

Simulating Wilson fermions, the bare mass in the Lagrangian undergoes
an additive renormalization, so that the chiral limit is reached for a
critical value $m_c$ which needs to be determined
nonperturbatively. Again perturbation theory is useful to get some
guidance on the initial choice of parameters before embarking in
actual simulations.

Following the notation in Ref.~\cite{Follana:2000mn}, we write the
perturbative expansion of the one--particle irreducible two point
function at zero momentum as:
\begin{equation}
m_c(g_0)=g_0^2 \Sigma^{(1)} + g_0^4 \Sigma^{(2)} + \ldots
\end{equation}
At one loop, the usual tadpole and sunset diagrams give rise to two
contributions, $c_1^{(1)}$ and $c_2^{(1)}$, respectively; for a
generic fermionic representation $R$, these yield:
\begin{eqnarray}
\Sigma^{(1)}=2 C_2(R) \left[c_1^{(1)}+c_2^{(1)}\right].
\end{eqnarray}
At this order in perturbation theory, the additive renormalization of
the mass is simply proportional to the quadratic Casimir of the
fermionic representation, while the proportionality constant
$c_1^{(1)}+c_2^{(1)}$ is independent of the representation. We can
therefore use the value in Ref.~\cite{Follana:2000mn}:
\begin{eqnarray}
\left[c_1^{(1)}+c_2^{(1)}\right]=-0.162857058711(2).
\label{eq:c1c2}
\end{eqnarray}

At two loops we need to inspect the structure of the diagrams listed
in Fig.~2 of Ref.~\cite{Follana:2000mn}, and modify the group
theoretical factors to take into account the fact that the generators
appearing in the vertices involving both fermions {\it and}\ gluons
are in a generic representation, whereas generators inside
3- and 4-gluon vertices are still in the fundamental representation. 
For our present purposes, the result of Ref.~\cite{Follana:2000mn} 
(Eq. (9) therein) for $\Sigma^{(2)}$ can be recast in the form:
\begin{eqnarray}
\Sigma^{(2)}&=& C_2(F) \, N \, d_1 + C_2(F) \, n_f \, d_2 + C_2(F)^2 (d_3+d_4),\\
d_1&=& -0.001940(6) \qquad d_2 = 0.00237236(16)\\
d_3&=& -0.081429(8) \qquad d_4 = 0.01516325(12)
\end{eqnarray}
($C_2(F)$ is the quadratic Casimir operator in the fundamental
representation). In the above, contributions proportional to
$C_2(F)^2$ have been separated into $d_3$ (arising solely
from the two diagrams with a tadpole made out of the 4-gluon vertex)
and $d_4$ (coming from the remaining diagrams). The extension to an
arbitrary representation $R$ is now immediate:
\begin{eqnarray}
\Sigma^{(2)}&=& C_2(R) \, N \, d_1 + 2 C_2(R) \, T_R \, n_f \, d_2 +
C_2(R) \, C_2(F) \, d_3 + C_2(R)^2 d_4 \,,
\end{eqnarray}
where the quantities $d_i$ are left unchanged.

The prediction from perturbation theory can be improved by resumming a
specific infinite class of gauge invariant tadpole diagrams. This
method is known under the 
name of {\it cactus}\ dressing~\cite{Panagopoulos:1998xf}, and it has
been shown to provide improved estimates for various quantities of
interest, bringing them closer to nonperturbative results. Unlike
other approaches for improving perturbation theory, such as
Refs.~\cite{Parisi:1980pe,Lepage:1992xa}, 
this approach does not rely on any Monte Carlo data as input, and it
is therefore ideally suited for an exploratory study, such as the
present one. Cactus resummation for the one-loop result of the
critical mass simply amounts to dividing the result by a factor
$\tilde c_0$ (denoted $(1-w(g_0))$ in
Ref.~\cite{Panagopoulos:1998xf}), which is
independent of the fermion representation 
(since it arises from an all-order resummation of gluon diagrams), but
depends on $N$ and on the bare coupling constant $g_0$. The
factor $\tilde c_0$ is the solution of the following equation:
\begin{equation}
u \, e^{-u (N{-}1)/(2N)} \,\left[ \frac{N{-}1}{N} \, L^1_{N{-}1}(u) +2
L^2_{N{-}2}(u) \right] = \frac{g_0^2\,(N^2{-}1)}{4}, \qquad
\tilde c_0 \equiv \frac{g_0^2}{4 u}
\label{eq:ctilde0}
\end{equation}
($L^\alpha_\beta\,$ are Laguerre polynomials). For $N=2$ and $N=3$,
Eq.~(\ref{eq:ctilde0}) simplifies to:
\begin{equation}\begin{array}{lcl}
\tilde c_0 = e^{\displaystyle -g_0^2/(16\,\tilde c_0)}\,
\left(1-\frac{\displaystyle g_0^2}{\displaystyle 24\,\tilde c_0}\right),&\quad&(N=2)\\
\tilde c_0 = e^{\displaystyle -g_0^2/(12\,\tilde c_0)}\,
\left(1-\frac{\displaystyle g_0^2}{\displaystyle 8\,\tilde c_0}+
\frac{\displaystyle g_0^4}{\displaystyle 384\,\tilde
  c_0^2}\right),&\quad &(N=3).
\end{array}\end{equation}
Figure~\ref{fig:ctilde0} presents 
$\tilde c_0$~\cite{constantinou-2006-74}
as a function of $g_0^2$, for $N=2$ and
$N=3$. The range of $g_0^2$ values, for which a solution exists, 
extends from $g_0^2=0$ (where $\tilde c_0 = 1$) up to $16/\sqrt{9e}
\simeq 3.23$ ($N=2$) and $1.558$ ($N=3$); this covers the whole
region of physical interest. 
\FIGURE[ht]{
\includegraphics[width=0.8\textwidth]{FIGS/ctilde0_vs_g0squared.eps}
\caption{$\tilde c_0$ as a function of $g_0^2$ for 
  $N{=}2$ (red line) and $N{=}3$ (green line). }
\label{fig:ctilde0}
}
The one--loop resummed result thus simply yields:
\begin{equation}
m_c= g_0^2 2 C_2(R) \left[c_1^{(1)}+c_2^{(1)}\right] / \tilde c_0 \, ,
\label{eq:mc}
\end{equation}
where $[c_1^{(1)}+c_2^{(1)}]$ is given in Eq.~(\ref{eq:c1c2}) and $\tilde
c_0$ may be read off Fig.~\ref{fig:ctilde0}.

In actual numerical simulations, the hopping parameter $\kappa$ is
used instead of the bare mass $m_0$. Based on the studies in
Ref.~\cite{Follana:2000mn}, we decide to use the one--loop resummed
result as our estimate of the location of the massless theory in the
space of bare parameters. The critical value of
the hopping parameter is given by: $\kappa_c=1/(2 m_c +8)$, where
$m_c$ is as in Eq.~(\ref{eq:mc}).

In order to compute the decay constant of pseudoscalar mesons, the
renormalization of the axial current $Z_A$ is needed. While a
nonperturbative determination of the renormalization constant is
desirable, for the first exploratory studies we shall again rely on
perturbation theory to determine $Z_A$. The accuracy of perturbation
theory can be estimated by comparing the results for QCD, where both
a perturbative and a nonperturbative computations are available. 

The perturbative renormalization for the axial current with Wilson
fermions was originally computed in Ref.~\cite{Martinelli:1982mw} --
see also Ref.~\cite{Eichten:1989kb} for a useful collection of results
for the numerical integrals that appear in lattice calculations. The
computation of the renormalization constant for fermion bilinears
requires the computation of the vertex functions, and of the fermion
wave--function renormalization. For both quantities the Feynman
diagrams that appear at one loop depend on the quadratic Casimir of
the fermion representation. They are therefore readily generalized to
an arbitrary representation, e.g.
\begin{eqnarray}
  \label{eq:ZV}
  Z_V&=&1 -\frac{g_0^2}{16\pi^2} C_2(R) 20.62, \\
  \label{eq:ZA}
  Z_A&=&1 - \frac{g_0^2}{16\pi^2} C_2(R) 15.7,
\end{eqnarray}
where the numerical factor is determined by numerical integrals that
do not depend on the fermion representation. Using the values for
$C_2(R)$ in the appendix, the values of $Z_A$ and $Z_V$ can be easily
computed for a generic representation.

The one-loop perturbative results for $Z_V,\ Z_A$ can also be improved
via cactus dressing; such an improvement has been known to work
rather well with fermions in the fundamental
representation~\cite{panagopoulos-1999-59}. 
To this effect, all that is required is a substitution of $g_0^2$ by
$g_0^2/\tilde c_0$ in Eqs.~(\ref{eq:ZV},\ref{eq:ZA}), where $\tilde c_0$
is again read off Fig.~\ref{fig:ctilde0}. 

\section{Generalized Aoki phases}
\label{sec:d}

The low--energy dynamics of the pseudo Goldstone particles is
determined by the pattern of chiral symmetry breaking.  For theories
with $n_f>1$ and fermions in a complex representation, such pattern is
the same as for QCD, and therefore we expect the same Lagrangian to
define the dynamics of the effective theory; clearly the low--energy
constants that appear in the Lagrangian do depend on the specific
model under study. In this section we discuss the general structure of
the chiral Lagrangian for fermions in arbitrary representations, and
the possible phases of the theories discretized on the lattice.

\subsection{$\mathop{\rm SU}(2)\times \mathop{\rm SU}(2) \rightarrow
  \mathop{\rm SU}(2)$}
For the theory T1 that we have been considering in this work, we have
two flavours in a complex representation of the gauge group and
therefore the usual $\mathrm{SU}(2)\times \mathop{\rm SU}(2)$ chiral
Lagrangian is expected to determine the dynamics in the low--energy
theory. Following the notation used in Ref.~\cite{Sharpe:1998xm}, and
including the symmetry breaking terms, yields the Lagrangian:
\begin{equation}
  \label{eq:chlagr}
  {\mathcal L}=\frac{F^2}{4} \mathrm{Tr}
  \left( \partial^\mu \Sigma^\dagger \partial_\mu \Sigma \right) 
  + \frac{c_1}{4} \mathrm{Tr} \left(\Sigma + \Sigma^\dagger\right)
  - \frac{c_2}{16} \left\{\mathrm{Tr} \left(\Sigma + \Sigma^\dagger\right)\right\}^2.
\end{equation}
The pion decay constant in technicolour theories is related to the
vacum expectation value of the (composite) Higgs field, which yields
$F\approx 250~\mathrm{GeV}$.  Lattice artefacts for Wilson fermions
enter in the coefficients of the symmetry breaking terms:
\begin{equation}
  \label{eq:latart}
  c_1 \sim \Lambda^3 \left(m + a\Lambda^2\right),~~~~~
  c_2 \sim \Lambda^2 \left(m^2 + m a\Lambda^2 + a^2 \Lambda^4\right),
\end{equation}
where $\Lambda$ is again the hadronic scale for the theory under
consideration.  The pattern of symmetry breaking depends on the
coefficients $c_1$ and $c_2$. Since these coefficients depend on the
PCAC mass $m$ and the lattice spacing $a$, the phase diagram can be
mapped into the plane of the bare parameters $m_0,g_0$, used in
lattice simulations. 

The analysis in Ref.~\cite{Sharpe:1998xm} remains unchanged; for the
theory with two flavours the field $\Sigma$ can be parametrized as:
\begin{equation}
  \label{eq:sigmapar}
  \Sigma=A + i {\mathbf B} \cdot {\mathbf \sigma},~~~~~A^2+{\mathbf B}^2=1,
\end{equation}
and the potential becomes:
\begin{equation}
  \label{eq:chipot}
  -c_1 A + c_2 A^2,
\end{equation}
so that the minimum of the potential $\Sigma_0=A_0 + i {\mathbf B}_0
\cdot {\mathbf \sigma}$ can develop a non--trivial ${\mathbf B}_0$
only if $|A_0|<1$. For $c_2>0$ a region of width $\sim
\left(a\Lambda\right)^3$ may exist, where the minimum of the potential
leads to an Aoki phase. Hence the approach to the chiral limit in
theories with two Wilson fermions in any complex representation is
similar to the one observed in QCD: as the quark mass is reduced at
fixed lattice spacing, flavour symmetry is broken and two massless
Goldstone bosons appear. The actual values of $c_1$ and $c_2$ depend
on the dynamics of the theory under study, and need to be estimated
for the cases of interest. Nevertheless, in all cases, the chiral
limit is entangled with the continuum limit, and the quark mass cannot
be lowered to arbitrarily small masses at fixed lattice spacing.

\subsection{$\mathop{\rm SU}(4)\rightarrow \mathop{\rm SO}(4)$}
In considering theories in arbitrary representations, different
patterns of chiral symmetry breaking may occur. The symmetry breaking
patterns and the effective theories describing the low--energy
dynamics in these cases have been studied e.g. in
Refs.~\cite{Peskin:1980gc,Appelquist:1999dq,Basile:2004wa,Foadi:2007ue}. Using
this effective theory framework, we discuss the possibility of having
an Aoki phase in one case that arises in phenomenologically
interesting theories, namely the breaking pattern SU(4)$\to$SO(4).

This symmetry breaking pattern appears for two Dirac fermions in the
adjoint representation of the gauge group, see e.g.  the theory T2
discussed above. Denoting by $\Sigma$ the Goldstone matrix, the
relevant effective potential for the study of the Aoki phase is:
\begin{eqnarray}
  V = - \frac{c_1}{4}{\rm Tr} \left[\Sigma + \Sigma^{\dagger}\right]  
  + \frac{c_2}{16} \left\{\mathrm{Tr} \left(\Sigma + \Sigma^\dagger\right)\right\}^2.
\end{eqnarray}
Here $\Sigma$ transforms linearly under the global symmetry group
$\mathrm{SU}(4)$, i.e.
\begin{equation}
\Sigma \rightarrow g \Sigma g^T  \quad {\rm with} \quad g\in \mathrm{SU}(4) \ ,
\end{equation}
and 
\begin{eqnarray}
\Sigma=\Sigma_0\,\exp (i \sum_a^9\frac{{\pi}^a}{f_\pi} X^a)  \ .
\end{eqnarray}
In discussing the possibility of an Aoki phase, we are interested in
finding at least {\it one}\ vacuum configuration where the condensate
is not proportional to the identity.  Indicating with $X^a$ and
$a=1,\ldots, 9$ the generators spanning the $\mathop{\rm
  SU}(4)/\mathop{\rm  SO}(4)$ quotient
space~\cite{Peskin:1980gc,Foadi:2007ue} (an explicit representation of
the matrices $X^a$ is reported in the appendix), we look for solutions
of the form:
\begin{eqnarray}
\Sigma_0 =2\sqrt{2}( A_0 + i\,\sum_a^9A^a X^a) \ ,  
\end{eqnarray}
where $A^a$ are real coefficients.  With this Ansatz, and using the
explicit expressions for the generators, the unitarity constraint,
i.e. $\Sigma_0^{\dagger} \Sigma_0 ={\mathbf 1}$, implies:
\begin{eqnarray}
  A_0^2 + \frac{( A_1^2 + A_2^2+A_3^2) (A_3^2+A_9^2 + A_8^2)}{A_3^2}=1 \ .
\end{eqnarray}
Substituting the expression for $\Sigma_0$ in the potential we find:
\begin{eqnarray}
  V = 4A_0^2 \, c_2  - 2 c_1 A_0  \ .
\label{panz}
\end{eqnarray}
For certain values of the potential coefficients $c_1,c_2$, there
exists a minimum for $A_0$ smaller than one. For instance, by taking
$A_1 = A_2=A_8=A_9=0$, but a nonzero $A_3$, we can satisfy the
unitarity constraints and $\mathop{\rm SO}(4)$ breaks spontaneously to
$\mathop{\rm {}U}(1)\times \mathop{\rm {}U}(1)$. That this is the
correct symmetry of the vacuum can be checked by determining which
$\mathop{\rm SO}(4)$ generator commutes with $X^3$.  We have checked
that the $S^3$ and $S^4$ generators of $\mathop{\rm SO}(4)$ explicitly
constructed in Ref.~\cite{Foadi:2007ue} are left unbroken and
constitute the generators of $\mathop{\rm {}U}(1)\times \mathop{\rm
  {}U}(1)$. In this case we would expect the emergence of an Aoki
phase with four Goldstone bosons associated to the breaking of
$\mathop{\rm SO}(4)$ to $\mathop{\rm {}U}(1)\times \mathop{\rm
  {}U}(1)$.

\subsection{$\mathop{\rm SU}(4)\rightarrow \mathop{\rm Sp}(4)$}
This symmetry breaking pattern does appear for fermions in
pseudo--real representations. The analysis is similar to the one done
before, now using the five (rather than nine) $X$ generators presented
in the appendix of Ref.~\cite{Appelquist:1999dq}. We seek for a
solution using the same Ansatz as in the preceding subsection. The
potential evaluated on the Ansatz is identical to the one in
Eq.~(\ref{panz}). In this case the unitarity constraint for $\Sigma_0$
yields the condition:
\begin{eqnarray}
A_0^2 + A_1^2+A_2^2+A_3^2+A_4^2+A_5^2 = 1 \ ,
\end{eqnarray}
For $A_0=1$ one recovers the $\mathop{\rm Sp}(4)$ symmetry. On the
other hand a minimum with $A_0<1$ implies that the $\mathop{\rm
  Sp}(4)$ symmetry is spontaneously broken to $\mathop{\rm SO}(4)$. We
hence have four broken generators of $\mathop{\rm Sp}(4)$
corresponding to $S^1,S^2,S^9$ and $S^{10}$ in the appendix of
Ref.~\cite{Appelquist:1999dq}. Again we find that an Aoki phase is
possible, with four massless bosons.
 
\subsection{Eigenvalues of the Dirac operator}
Finally let us comment briefly on the role of the small eigenvalues of
the Dirac operator, since they play a crucial role in theories where
chiral symmetry is spontaneously broken.  It was indeed realized long
ago that in QCD the chiral condensate is related to the density of
eigenvalues of the Dirac operator. This property is encoded in the
Banks--Casher relation~\cite{Banks:1979yr}:
\begin{equation}
  \label{eq:bankscasher}
  \langle0|\bar q q|0\rangle=-\pi \rho(0),
\end{equation}
where $\rho(\lambda)$ is the number of eigenvalues in the interval
$d\lambda$ per unit volume. 

Following e.g. the derivation in Ref.~\cite{Leutwyler:1992yt}, it is
straightforward to show that the same relation holds independently of
the fermionic representation. As a consequence, we expect to have a
finite density of eigenvalues around $\lambda=0$ for any theory that
breaks chiral symmetry spontaneously developing a non--zero chiral
condensate, and hence the number of small eigenvalues of the Dirac
operator grows with the four--dimensional volume as the mass tends to
zero. This phenomenon is directly related to the spontaneous breaking
of chiral symmetry, and does not depend on the particular lattice
discretization.

Moreover, for lattice formulations that do not preserve chiral
symmetry, such as the Wilson formulation, the spectrum of the
hermitian Dirac operator is not bounded from below by the bare quark
mass. As the quark mass is lowered at fixed lattice spacing, the
probability of finding an exceptionally small eigenvalue becomes
non--negligible. These small eigenvalues lead to algorithmic
instabilities, violations of ergodicity, and sampling inefficiencies,
which could seriously distort the output of numerical
simulations~\cite{DelDebbio:2005qa}. A careful study of the spectrum
of the Dirac operator for theories in higher representations is
therefore necessary in order to determine a safe region for simulating
Wilson fermions. This is particularly important as one tries to study
the phase diagram of novel and unknown theories.

\section{Conclusions}
Gauge theories with fermions in higher--dimensional representations
have been put forward in several contexts; they are important both for
phenomenological and theoretical studies. Some of them provide viable
candidates for strong electroweak symmetry breaking, that are not
ruled out by precision measurements. On a more theoretical side, they
''interpolate'' between supersymmetric and non--supersymmetric
theories, thereby opening new ways to try to tame the nonperturbative
dynamics of gauge theories. Due to the recent progress in numerical
simulations of gauge theories on the lattice, it is now possible to
simulate these theories, and some first works in this direction have
already appeared. In this work, we have generalized known analytical
results to the case of fermions in arbitrary representations. In
particular, we have considered the scaling as the continuum limit is
approached, the location of the critical bare hopping parameter
corresponding to massless quarks, the renormalization of fermionic
bilinears, and the phase structure as the quark mass is lowered at
fixed lattice spacing. The results presented here provide some insight
on the unknown phase diagram of these theories and will be useful to
guide forthcoming simulations. Definitive answers on the strong
dynamics of such theories, and therefore about their viability as
phenomenological candidates, can only be provided by actual numerical
simulations.

\acknowledgments LDD is supported by an STFC Advanced Fellowship. LDD
would like to thank the Isaac Newton Institute for hospitality while
this work was completed. The work of M.T.F. and F.S. is supported by
the Marie Curie Excellence Grant under contract
MEXT-CT-2004-013510. Some of the ideas in this paper were discussed
during the workshop {\it Strongly interacting dynamics beyond the
  Standard Model}, for which we acknowledge financial support from
SUPA.

\appendix

\section{Group--theoretical factors}
\label{app:a}
The normalization of the generators in a generic representation $R$
of $\mathrm{SU}(N)$ is fixed by requiring that:
\begin{equation}
[T^a_R, T^b_R] = {\rm i}\, f^{abc} T^c_R\,,
\end{equation}
where the structure constants $f^{abc}$ are the same in all representations.
We define:
\begin{eqnarray}
\mathrm{tr }_R \left(T^a T^b \right) = \mathrm{tr } \left(T^a_R T^b_R
\right) &=& T_R \delta^{ab}, \\
\sum_a \left(T^a_R T^a_R \right)_{AB} &=& C_2(R) \delta_{AB},
\end{eqnarray}
and hence: 
\begin{equation}
T_R = \frac{1}{N^2-1} C_2(R) d_R
\end{equation}
where $d_R$ is the dimension of the representation $R$.  The quadratic
Casimir operators may be computed from the Young tableaux of the
representation of $\mathrm{SU}(N)$ by using the formula:
\begin{equation}
C_2(R) =\frac{1}{2}\left(nN+ \sum_{i=1}^{m} n_i \left( n_i+1-2i
\right) - \frac{n^2}{N}\right)
\end{equation}
where $n$ is the number of boxes in the diagram, $i$ ranges over the
rows of the Young tableau, $m$ is the number of rows, and $n_i$ is the
number of boxes in the $i$-th row. 

\TABLE[ht]{
\caption{Group invariants used in this work}
\label{table1}
\begin{tabular}{r|c|c|c}
R    & $d_R$               & $T_R$           & $C_2(R)$            \\
\hline
fund & $N$                 & $1/2$       & $(N^2-1)/(2 N)$ \\        
Adj  & $N^2-1$             & $N$             & $N$ \\
2S   & $N(N+1)/2$ & $(N+2)/2$ & $C_2(F) 2(N+2)/(N+1)$ \\
2AS  & $N(N-1)/2$ & $(N-2)/2$ & $C_2(F)
\frac{2(N-2)}{N-1}$ \\
\hline
\end{tabular}
}

The quantities $d_R$, $T_R$, $C_2(R)$ are listed in Table~\ref{table1}
for the fundamental, adjoint, 2--index symmetric, and 2--index
antisymmetric representations.

The generators $X^a$ spanning the $\mathrm{SU}(4)/\mathrm{SO}(4)$
quotient space are defined as:
\begin{equation} X^i = \frac{1}{2\sqrt{2}}\begin{pmatrix} \tau^i & 0 \\
0 & (\tau^i)^T \end{pmatrix}, \ 1\le i \le 3;
\quad
X^i = \frac{1}{2\sqrt{2}}\begin{pmatrix} 0 & D^i \\
(D^i)^\dag & 0 \end{pmatrix}, \  4\le i \le 9,
\end{equation}
with
\begin{equation}\begin{array}{r@{\;}c@{\;}lr@{\;}c@{\;}lr@{\;}c@{\;}l}
D^4 &=& \mathbbm{1} \ , & \quad D^6 &=& \tau^3 \ , & \quad D^8 &=& \tau^1 \ , \\
D^5 &=& i\mathbbm{1} \ , & \quad D^7 &=& i\tau^3 \ , & \quad D^9 &=& i\tau^1
\ .
\end{array}\end{equation}

%\bibliography{hirep}
\providecommand{\href}[2]{#2}\begingroup\raggedright\endgroup

\end{document}